\documentclass[nofootinbib,aps,prl,preprint,longbibliography]{revtex4-1}
\usepackage{amsmath}
\usepackage{bbold}
\usepackage{graphicx}
\usepackage[pdfencoding=auto]{hyperref}
\usepackage{bookmark}
\usepackage{lineno}
\newcommand{\parallelsum}{\mathbin{\!/\mkern-5mu/\!}}

\begin{document}
\title{Nanometric skyrmion lattice from anisotropic exchange interactions in a centrosymmetric host }
\author{Max Hirschberger$^{1,2}$}\email{hirschberger@ap.t.u-tokyo.ac.jp}
\author{Satoru Hayami$^{1}$}
\author{Yoshinori Tokura$^{1,2,3}$}
\date{\today}
\affiliation{$^{1}$Department of Applied Physics and Quantum-Phase Electronics Center, The University of Tokyo, Bunkyo-ku, Tokyo 113-8656, Japan}
\affiliation{$^{2}$RIKEN Center for Emergent Matter Science (CEMS), Wako, Saitama 351-0198, Japan}
\affiliation{$^{3}$Tokyo College, The University of Tokyo, Bunkyo-ku, Tokyo 113-8656, Japan}

\maketitle

\begin{center}
\begin{large}
Abstract
\end{large}
\end{center}
Skyrmion formation in centrosymmetric magnets without Dzyaloshinskii-Moriya interactions was originally predicted from unbiased numerical techniques. However, no attempt has yet been made, by comparison to a real material, to determine the salient interaction terms and model parameters driving spin-vortex formation. We identify a Hamiltonian with anisotropic exchange couplings, local ion anisotropy, and four-spin interactions, which is generally applicable to this class of compounds. In the representative system Gd$_3$Ru$_4$Al$_{12}$, anisotropic exchange drives a fragile balance between helical, skyrmion lattice, and transverse conical (cycloidal) orders. The model is severely constrained by the experimentally observed collapse of the SkL with a small in-plane magnetic field. For the zero-field helical state, we further anticipate that spins can be easily rotated out of the spiral plane by a tilted magnetic field or applied current.

\section{Introduction}
The modern push to realize complex magnetism with strong coupling to the electronic degrees of freedom is underpinned, to a large extent, by ever more powerful computational techniques. Amongst these are unbiased numerical simulations, such as Monte-Carlo methods, for magnetic ground states, excitation spectra, and spin dynamics starting from an (effective) spin-spin Hamiltonian. Monte-Carlo studies originally predicted that centrosymmetric triangular lattice magnets, be they insulators with (superexchange) interactions \cite{Okubo2012,Leonov2015} or metals with effective Ruderman-Kittel-Kasuya-Yosida (RKKY) couplings \cite{Ozawa2017,Hayami2017,Wang2020}, can host nanometer-sized magnetic spin vortices with net topological charge (i.e., skyrmions), and a multitude of even more complex states \cite{Ozawa2017,Shimokawa2019, Lohani2019}. The theoretical work has introduced a new playground for the application of concepts from topology in condensed matter, but only with the observation of skyrmion lattices in the hexagonal intermetallics Gd$_2$PdSi$_3$ \cite{Kurumaji2019} and Gd$_3$Ru$_4$Al$_{12}$ \cite{Hirschberger2019} has a more systematic comparison of theory and experiment become possible.

In this article, we demonstrate via a combination of theory and experiments that moderately weak anisotropic exchange from dipolar and spin-orbit interactions is sufficient to realize the skyrmion lattice state in centrosymmetric magnets. Anisotropic exchange hence represents a new, third avenue towards the SkL in bulk crystals with inversion center, beyond frustrated exchange and higher-order RKKY interactions. Next to the present focus on intermetallic systems, this notion may have implications also for spin-vortex formation in insulating materials, e.g. on the magnetic diamond lattice of MnSc$_2$S$_4$~\cite{Gao2017,Gao2020}. In noncentrosymmetric magnets and at interfaces, spin-orbit coupling (SOC) provides the crucial underpinning of spin-spiral and skyrmion formation via the Dzyaloshinskii-Moriya interaction~\cite{Dzyaloshinskii1957,Moriya1960,Bak1980,Roessler2006}. Likewise, we here show that SOC can play an important role in centrosymmetric magnets with SkL, as a driver of anisotropic exchange. Our work on weakly anisotropic exchange also creates a conceptual link between SkL formation and the movement to realize new quantum-disordered states in frustrated magnets via strongly bond-anisotropic exchange~\cite{Kitaev2006,Jackeli2009}.

The focus of our study is Gd$_3$Ru$_4$Al$_{12}$, a representative of the new class of centrosymmetric skyrmion hosts with coupled local moments and itinerant electrons. The Gd$^{3+}$ magnetic moments in this structure are arranged in quasi-layered Kagome nets, which are distorted by alternate stretching and compression of Kagome bond distances~\cite{Gladyshevskii1993,Niermann2002,Chandragiri2016} – termed breathing Kagome lattice [Fig. \ref{fig:fig1}(a)]. The magnetic phase diagram for field along the crystallographic $c$-axis (perpendicular to the Kagome plane) harbors five distinct regimes \cite{Hirschberger2019,Matsumura2019}: Helical order (HE), a Transverse Conical state (TC), fan-like order (F), the skyrmion lattice (SkL), and the as-yet uncharacterized pocket labeled by the Roman numeral $V$ [Fig. \ref{fig:fig1}(b)]. In our analysis, a key guiding factor for the numerical simulation is the stability range of the SkL as compared to TC.

\section{Model Hamiltonian}
\begin{figure}[!htb]
  \begin{center}
		\includegraphics[clip, trim=0.5cm 0.0cm 0.5cm 0.cm, width=1.\linewidth]{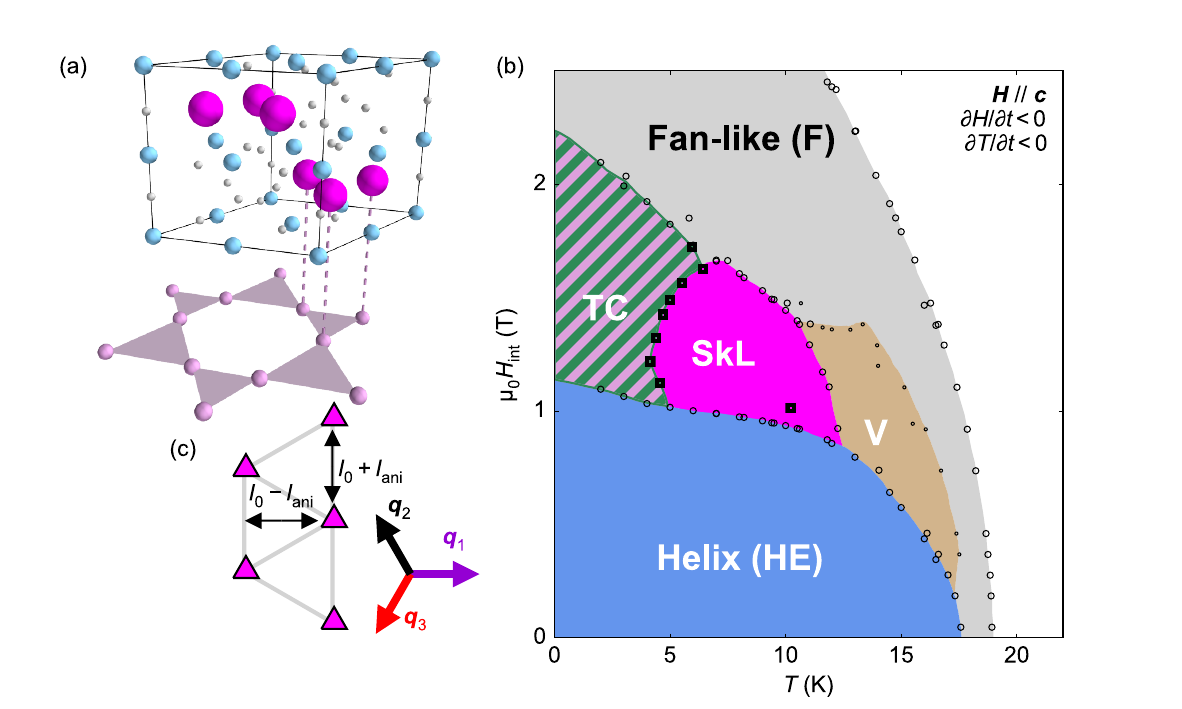}
		\internallinenumbers
    \caption[]{(color online). Magnetic interactions in centrosymmetric Gd$_3$Ru$_4$Al$_{12}$. (a) Hexagonal crystal structure with Al, Ru, and Gd atoms marked by white, blue, and magenta spheres, respectively. (b) Magnetic phase diagram for magnetic field applied along the $c$-axis. (c) The distorted Kagome motif of rare earth sites is approximated as a triangular lattice of trimer plaquettes (magenta triangles). Three possible directions, equivalent by symmetry, of the magnetic ordering vector $\mathbf{q}_\nu$ ($\nu = 1,2,3$) are indicated in the figure. Specifically for $\mathbf{q}_1$, we illustrate the anisotropy of parallel and perpendicular magnetic exchange interactions by black arrows. Data points in (b) adopted from Ref. \cite{Hirschberger2019}.}
    \label{fig:fig1}
  \end{center}
\end{figure}

Motivated by the structural feature of rare earth triangles, and by previous considerations of the specific heat \cite{Nakamura2018}, we approximately treat Gd$_3$Ru$_4$Al$_{12}$ as a two-dimensional triangular lattice of strongly coupled superspin trimers [small triangles in the projection plane of Fig. \ref{fig:fig1}(a)]. Figure \ref{fig:fig1}(b) shows that the zero-field magnetic ground state of hexagonal Gd$_3$Ru$_4$Al$_{12}$ is a spiral, where the magnetic modulation vector $\mathbf{q}$ is aligned within the basal plane. In fact, three $\mathbf{q}_\nu$ with $\nu=1,2,3$ must be degenerate by symmetry, and previous work has shown that the $\mathbf{q}_\nu$ point along the $a^*$ axis and equivalent directions \cite{Hirschberger2019,Matsumura2019}. Backed by the knowledge that the directors $\hat{\mathbf{q}}_\nu$ are identical in all phases experimentally studied so far, we choose a minimal Hamiltonian consistent with the six-fold symmetry of the lattice and suitable for discussion of the magnetic phase transitions \cite{Li2016}. Only the dominant Fourier components $\mathcal{S}_{\mathbf{q}_\nu}^\alpha$ of the local spin $\pmb{\mathcal{S}}_i$ at lattice site $i$ are carried over \cite{Hayami2018}, so that
\begin{align}
\mathcal{H}/J &=2\sum_\nu \left[-\mathcal{H}_\nu+(K/N)\left(\mathcal{H}_\nu\right)^2 \right] \notag \\
&-A\sum_i \left(\mathcal{S}_i^z \right)^2 -h\sum_i \hat{\mathbf{h}} \cdot \pmb{\mathcal{S}}_i \label{eq:hamil}
\end{align}
with $\mathcal{H}_\nu = \sum_{\alpha,\beta}\Gamma_{\mathbf{q}_\nu}^{\alpha\beta}  \mathcal{S}_{\mathbf{q}_\nu}^\alpha \mathcal{S}_{-\mathbf{q}_\nu}^\beta$. Here, $J$ is the energy scale for the dominant two-spin interaction, while ($K/N$), $A$, and $h$ are dimensionless parameters corresponding to the strength of four-spin interaction, local ion anisotropy, and external magnetic field, respectively (normalized by $J$ in all three cases). $N$ is the number of spins in the system, and the direction of the applied magnetic field is labeled as $\hat{\mathbf{h}}$. The matrices $\Gamma_{\mathbf{q}_\nu}^{\alpha\beta}$ are also dimensionless and characterize the anisotropic exchange interactions. For example for $\nu=1$, we have $\Gamma_{\mathbf{q}_1}=\text{diag}\left(I_0-I_\text{ani},I_0+I_\text{ani},I_z \right)$; the other $\Gamma_{\mathbf{q}_\nu}$ are obtained from $\Gamma_{\mathbf{q}_1}$ by a simple rotation operation of $\pm 120^\circ$. In contrast to models of magnetic interactions in real space, the present (momentum-space) approach for the two-spin and four-spin terms is not sensitive to the choice of boundary conditions \cite{Rybakov2016}. For simplicity of the numerical treatment, the calculations were carried out for $\mathbf{q}_\nu$ rotated by $90^\circ$ from those in Fig. \ref{fig:fig1}(c), i.e. for $\mathbf{q}_\nu$ along the triangular lattice's bond direction.

Anisotropic interactions of the $I_\text{ani}$-type can be caused by relativistic spin-orbit coupling \cite{Li2016}, but we note that dipolar interactions at the nearest neighbor level are also included in this term. In Discussion, we comment on the prospect of fully ab-initio calculations of $I_\mathrm{ani}$ in this class of compounds. More broadly speaking, Hamiltonians analogous to Eq. (\ref{eq:hamil}) are believed to be a good starting point as well for the description of phase transitions in centrosymmetric metallic magnets of different symmetries, such as tetragonal or cubic, if the $\mathbf{q}_\nu$ are strongly pinned to preferred crystal axes. Moreover, the formalism can be adapted to the description of non-centrosymmetric materials~\cite{Puphal2020} by inclusion of antisymmetric terms $\Gamma_{\mathbf{q}_\nu}^{\alpha \beta}\neq \Gamma_{\mathbf{q}_\nu}^{\beta\alpha}$, although this implies a larger number of adjustable parameters. When allowing for moderate ($\%$-scale) changes of the modulus $q_\nu=\left|\mathbf{q}_\nu \right|$ at phase boundaries –- including transitions from orders incommensurate with the underlying crystal lattice, to commensurate ones –- the energy landscape is modified but weakly due to the typically very broad magnetic susceptibility (Lindhard function) $\chi(\mathbf{q})$ in metallic magnets with large, localized magnetic moments \cite{Nomoto2020}. 

Without loss of generality, $I_z=I_0=1$ is used in the following, as small differences between $I_0$ and $I_z$ can be absorbed into $A$, where large (small) $I_z$ corresponds to positive (negative) $A$. Previous magnetization measurements indicated mild easy-plane anisotropy $A<0$ of local ions in Gd$_3$Ru$_4$Al$_{12}$ \cite{Nakamura2018,Hirschberger2019}. As $A<0$ favors the transverse conical state (TC) already in zero magnetic field, $I_\text{ani}>0$ is crucial to obtain phase HE with spiral vectors $\mathbf{q}_\nu\,\parallelsum\,a^*$ at $h=0$. In other words, stronger interactions parallel to the direction of the bond between triangular lattice sites are necessary to obtain phase HE in the present compound [Fig. \ref{fig:fig1}(c)].

Detailed supporting calculations show that excess $I_\mathrm{ani}$ strongly favors multi-$\mathbf{q}$ phases, promoting the SkL over TC at all temperatures and eventually causing an instability of HE towards a multi-$\mathbf{q}$ state \cite{SI}. To describe Gd$_3$Ru$_4$Al$_{12}$, we focus on $0<I_\text{ani}<0.1$ and set $K=0$, the latter choice being revisited below. Within this range of parameters, the phase boundaries at $\hat{\mathbf{h}}\,\parallelsum\, c$ are rather robust. For example, the simulations demonstrate that in absence of single ion anisotropy ($A=0$), the overall properties of the $c$-axis phase diagram are nearly unchanged as compared to $(I_\text{ani},A) = (0.01,-0.007)$. As we will show now, a good match of experiment and theory is obtained for $I_\text{ani}=0.01$ and $A= -0.007$, especially when aiming to also describe the case of $\hat{\mathbf{h}}$ tilted away from the $c$-axis.

\section{Unbiased modeling of magnetic phase transitions for $\hat{\mathbf{h}}\parallelsum\,c$}
\begin{figure*}[!htb]
  \begin{center}
		\includegraphics[clip, trim=0.6cm 6.5cm 1.0cm 3.0cm, width=1.\linewidth]{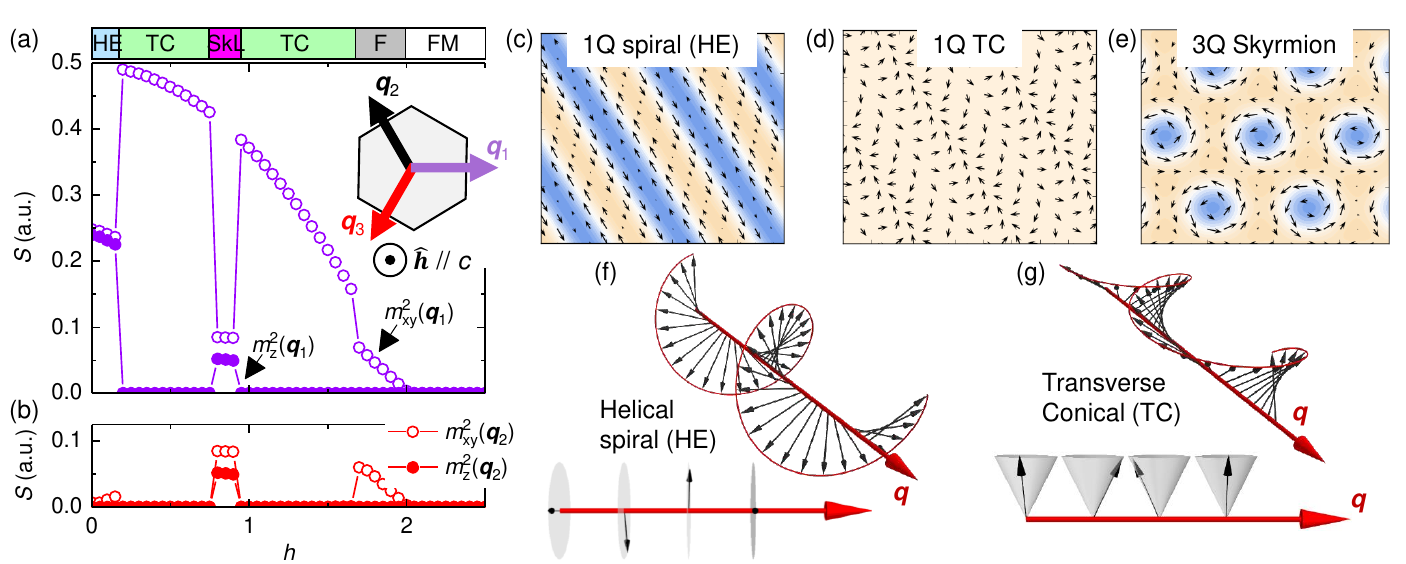}
		\internallinenumbers
    \caption[]{(color online). Simulated-annealing of magnetic order on the triangular lattice, for magnetic field $\hat{\mathbf{h}}\parallelsum\,c$. $h$ is the magnitude of the applied magnetic field measured in units of the exchange coupling $J$. (a,b), Components of the magnetic structure factor $S_\alpha (\mathbf{q}_\nu)\sim m_\alpha^2 (\mathbf{q}_\nu)$, where $m_\alpha (\mathbf{q}_\nu )$ is a vector component of the modulated magnetic moment at $\mathbf{q}_\nu$ with $\nu=1,2$. The simulated annealing calculation is carried out~\cite{SI} for parameters $I_\text{ani}=0.01$, $A=-0.007$ at low temperature $T=0.01\cdot J$. Inset of (a), the three-fold degeneracy of the $\mathbf{q}_\nu$ in the hexagonal plane. Colored bands at the top define parameter regimes for phases HE (helical), TC (transverse conical), SkL (skyrmion lattice), F (fan-like), and field-aligned ferromagnet (FM). In (c-e) magnetic textures are shown by arrows (magnetic moment in the hexagonal  plane) and the color map, where yellow and blue correspond to positive and negative $c$-component of the moment, respectively. In (f,g) we depict both a three-dimensional sketch of the orders in phases HE and TC, and an illustration in terms of spiral planes and cones (lower side).}
    \label{fig:fig2}
  \end{center}
\end{figure*}

Figure \ref{fig:fig2}(a) shows the components $S_\alpha (\mathbf{q}_\nu)$ of the calculated magnetic structure factor, which is proportional to the squared Fourier component of the magnetic moment $m_\alpha (\mathbf{q}_\nu )^2$. We use $S_\alpha (\mathbf{q}_\nu)$ and $m_\alpha (\mathbf{q}_\nu )^2$ interchangeably in the following. Here, the indices $\alpha =\parallel,\perp,z$ relate to a Cartesian frame of reference with unit vectors $\mathbf{e}_{\parallel \nu}=\hat{\mathbf{q}}_\nu$, $\mathbf{e}_{z\nu}=\hat{z}$, and $\mathbf{e}_{\perp \nu}$ perpendicular to both. For convenience, we also define $m_{xy}^2=m_\parallel^2+m_\perp^2$.  For the chosen model parameters, the zero-field ground state corresponds to a helical spiral (HE) with $\mathbf{q}=\mathbf{q}_1$ [Fig. \ref{fig:fig2}(d,g)] with very weak admixtures of fan-like components at $\mathbf{q}_2$, $\mathbf{q}_3$ \cite{SI}. When $\hat{\mathbf{h}}$ is applied parallel to the $c$-axis, the system transitions into a TC state (only $m_{xy}\neq 0$ for a single $\mathbf{q}_\nu$), then to the SkL ($m_{xy}\neq 0$, $m_z \neq 0$ with equal weight for all three $\mathbf{q}_\nu$), back into the TC phase, and finally into a high-field order which can be either transverse conical or fan-like, depending on the exchange parameters \cite{SI}. Sharp changes in $S_\alpha \left(\mathbf{q}_\nu \right)$ indicate first order phase transitions, which bound the SkL state and also characterize the spin-flop transition between HE and TC. Experimentally, these transitions were found to be accompanied by strong hysteresis. The strong history dependence of the experimental observations in phase TC, c.f. Fig. \ref{fig:fig1}(b) and Ref. \cite{Hirschberger2019}, in fact represents a challenge for the comparison to the theoretical result [Fig. \ref{fig:fig2}(a,b)]. We discuss agreement of the two phase diagrams in more detail in the Supplementary Information \cite{SI}.

\section{Experiment: Destruction of skyrmion lattice with in-plane magnetic field}
\begin{figure}[!htb]
  \begin{center}
		\includegraphics[clip, trim=0.cm 0.cm 0.cm 0.cm, width=1.\linewidth]{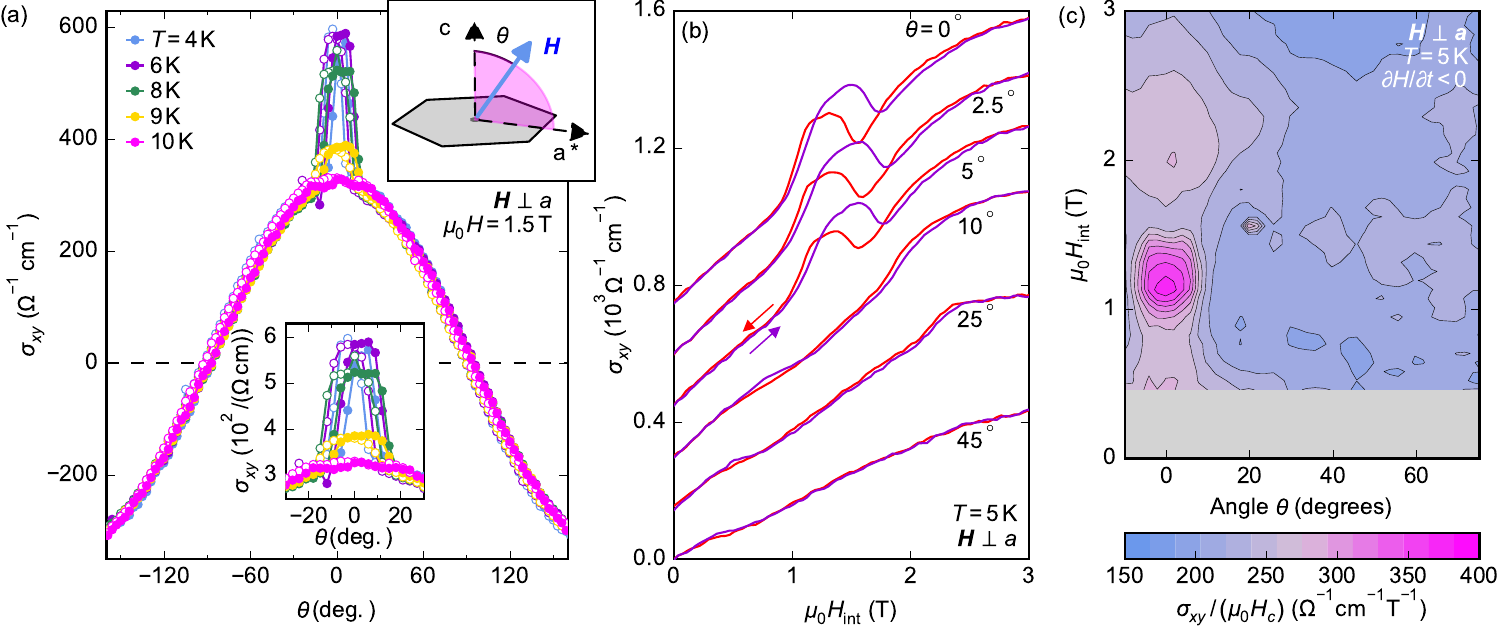}
		\internallinenumbers
    \caption[]{(color online). Collapse of skyrmion lattice (SkL) in Gd$_3$Ru$_4$Al$_{12}$ upon slight tilting of the magnetic field, as detected experimentally by the topological Hall effect. (a) Angle-dependent Hall conductivity $\sigma_{xy}$ at fixed magnetic field and temperature ($\mu_0 H=1.5\,$T). Open and solid symbols mark decreasing and increasing angle $\theta$. The right-hand and bottom insets illustrate the experimental geometry and show an expanded view of the data, respectively. (b) Field scan of $\sigma_{xy}$ at fixed $\theta$. Red and violet curves are for decreasing and increasing field, marked by the arrows. (c) The region of enhanced intensity in the contour map of $\sigma_{xy}/(\mu_0 H_c)$ is identified with the SkL phase. $H_c=H \cos\theta$ is the component of the magnetic field parallel to the $c$-axis. A demagnetization correction was applied to obtain the internal magnetic field $\mu_0 H_\text{int}$ for panels (b,c).}
    \label{fig:fig3}
  \end{center}
\end{figure}

The general characteristics of the calculated phase diagram for $\hat{\mathbf{h}}\,\parallelsum\,c$ are unchanged for a rather broad range of Hamiltonian parameters as long as $I_\text{ani}>0>A$ and $I_\text{ani}+A \ge 0$~\cite{SI}. To further constrain the two free parameters $A$ and $I_\text{ani}$, we experimentally studied the effect of tilting the magnetic field away from $c$ by an angle $\theta$, $c$ being the direction of six-fold symmetry (Fig. \ref{fig:fig3}). We hence establish that in Gd$_3$Ru$_4$Al$_{12}$, the critical angle $\theta_c$ of the skyrmion lattice phase is only $10-15^\circ$. In the experiment, we use the Hall conductivity $\sigma_{xy}$ as a sensor for the SkL phase. This observable is the transverse element of the conductivity tensor $\sigma_{\alpha\beta}$ relating electric field $E_\alpha$ and resulting charge current via $J_\alpha=\sigma_{\alpha\beta} E_\beta$. The Hall conductivity acquires an additional contribution in the SkL phase, the topological Hall effect (THE), which effectively measures the winding number of a single skyrmion \cite{Bruno2004,Neubauer2009,Ritz2013}. In our measurements of angle-dependent $\sigma_{xy}$, the THE emerges as a bell-shaped anomaly on top of a smooth background.

Outside the SkL phase, the curves of $\sigma_{xy} (\theta)$ collapse nicely onto a cosine-shaped profile. This behavior is consistent with the understanding that the oscillating background, on top of which the topological Hall signal $\sigma_{xy}^\mathrm{T}\approx 250\,\mathrm{\Omega^{-1}\,cm^{-1}}$ of the SkL develops, is due to spin-orbit coupling and the Karplus-Luttinger type anomalous Hall effect $\sigma_{xy}^\mathrm{KL}$ \cite{Nagaosa2010,Karplus1954}. For $\sigma_{xy}^\mathrm{KL}$, temperature dependence is not expected as long as the conduction electron’s spin polarization remains unchanged \cite{Lee2007}. Moreover, $\sigma_{xy}^\mathrm{KL}$ is proportional to the $c$-component of the net magnetization, consistent with the cosine-law observed here \cite{Nagaosa2010}.

The experimental $\sigma_{xy}$ data is reported as a function of the applied magnetic field $\mu_0 H$ in units of Tesla, where $\mu_0$ is the vacuum permeability. We corrected for the demagnetization effect by determining the internal field $\mu_0 H_\mathrm{int} = \mu_0 \left(H-NM\right)$, where $M$ is the magnetization per unit volume, measured independently, and $0\le N\le 1$ is the demagnetization factor calculated in elliptical approximation \cite{SI}. Given the presence of Gd$^{3+}$ ions with a moment of seven Bohr magneton ($\mu_B$) in this material, the corresponding dimensionless Zeeman energy is $h = 7\mu_B\cdot (\mu_0 H_\mathrm{int}) / J$~\cite{SI}.

The present experimental observations contrast previous findings on the far more stable skyrmion lattice with larger $\theta_c\sim 45^\circ$ in the related hexagonal compound Gd$_2$PdSi$_3$, with a triangular net of rare earth moments \cite{Kurumaji2019}. In Gd$_3$Ru$_4$Al$_{12}$, the delicate interplay between TC and SkL for $\mathbf{H}\parallelsum c$, and also the low value of $\theta_c$, are both consequences of a fine balance between the parameters $A$ and $I_\text{ani}$.

\section{Modeling the phase diagram in tilted field}
Let the in-plane component of $\hat{\mathbf{h}}$ be aligned along $\mathbf{q}_1$. We use the simulated annealing framework to first discuss the character of magnetic order above the critical angle $\theta_c$. Figures \ref{fig:fig4}(a,b) show how only a single modulation direction, $\mathbf{q}_1$, survives under these conditions, with spins arranged in a conical fashion around the axis $\hat{\mathbf{n}}$. In the simulation, $\hat{\mathbf{n}} \parallelsum\, \hat{\mathbf{h}}$ in this regime, so that the conical axis rotates smoothly as $\theta$ is changed. The resulting dependences $m_\parallel^2 \sim \cos^2 \theta$ and  $m_z^2\sim \sin^2 \theta$ are indicated by black solid lines in Fig. \ref{fig:fig4}(a). 

We are now in a position to explore the parameter range of $I_\text{ani}$ and $A$ suitable to describe the experimental situation in Gd$_3$Ru$_4$Al$_{12}$, when using the trimer approximation (Fig. \ref{fig:fig5}). The calculations demonstrate that for each $I_\text{ani}>0$ there is a critical $A_c$ which separates the stability regime of two zero-field ground states: For $A>A_c$, a helical spiral is the preferred spin configuration, but it succumbs to the transverse conical state when $A<A_c$. Figure \ref{fig:fig5} shows the $A_c$ values as dashed vertical lines for $I_\text{ani}=0.01,0.02,0.03$. The key point here is that only for small $I_\text{ani}=0.01-0.02$, the critical angle $\theta_c$ of the skyrmion lattice can be continuously tuned to zero before the collapse of the zero-field HE state. Following Fig. \ref{fig:fig5}, we conclude that the experimental situation in Gd$_3$Ru$_4$Al$_{12}$ is well described using the trimer approximation and $0<I_\text{ani}<0.02$ with correspondingly small, finely tuned $A$.

We return to a question which was postponed earlier: What is the role of the four-spin term $K$ in Eq. (1), or rather: could $K\neq 0$ explain the experiments for Gd$_3$Ru$_4$Al$_{12}$? $K$ is well known to strongly favor multi-$\mathbf{q}$ order such as the skyrmion lattice~\cite{Hayami2017}, and it comes as no surprise that $K>0$ enhances the critical angle $\theta_c$ for a variety of combinations of $(A, I_\mathrm{ani})$. For example when $I_\text{ani}=0.01$ and  $K=0.03$, the skyrmion lattice is stable even when $\hat{\mathbf{h}}$ is fully aligned along the in-plane direction $x$. As this is inconsistent with the observed small critical angle, we maintain that $K=0$ is suitable for the present breathing Kagome compound.

\section{Discussion}
Many previous numerical studies have relied on Heisenberg Hamiltonians which yield, in a certain parameter range, a SkL phase when aided by thermal fluctuations or easy-axis anisotropy $A>0$ \cite{Okubo2012,Leonov2015,Hayami2016}. These models are not well suited to the present material, where the SkL is realized despite $A<0$. Likewise, RKKY Hamiltonians with biquadratic interactions ($K\neq 0$) \cite{Ozawa2017,Hayami2017} are inconsistent, in our simulations, with the very small critical angle $\theta_c$ observed experimentally for Gd$_3$Ru$_4$Al$_{12}$. Hence, a new approach, emphasizing the role of anisotropic exchange $I_\text{ani}>0$ together with easy-plane anisotropy of rare earth ions, is used to model the stability ranges of helical, transverse conical, and SkL phases on the triangular lattice of superspin trimers, corresponding to the trimerized limit of the breathing Kagome network in Gd$_3$Ru$_4$Al$_{12}$~\cite{Nakamura2018,Matsumura2019}. Spin-orbit coupling \cite{Li2016} and next-neighbor dipolar interactions both contribute to $I_\text{ani}$. Note that easy-plane single-ion anisotropy was previously thought to be detrimental to skyrmion formation in magnets with inversion center \cite{Leonov2015,Hayami2016}. However, $A<0$ can stabilize Neel skyrmions in noncentrosymmetric polar materials \cite{Bordacs2017,Leonov2017}.

The prediction of $I_\mathrm{ani}$ for itinerant electron systems through fully ab-initio electronic structure calculations remains an open challenge in this field. $I_\mathrm{ani}$ is expected to be strongly dependent on the shape of Fermi surface sheets, as well as on the atomic spin-orbit coupling parameter $\lambda_\mathrm{SOC}$ of the orbitals constituting the conduction bands.

\begin{figure}[b]
  \begin{center}
		\includegraphics[clip, trim=1.cm 4.8cm 1.cm 3.2cm, width=0.95\linewidth]{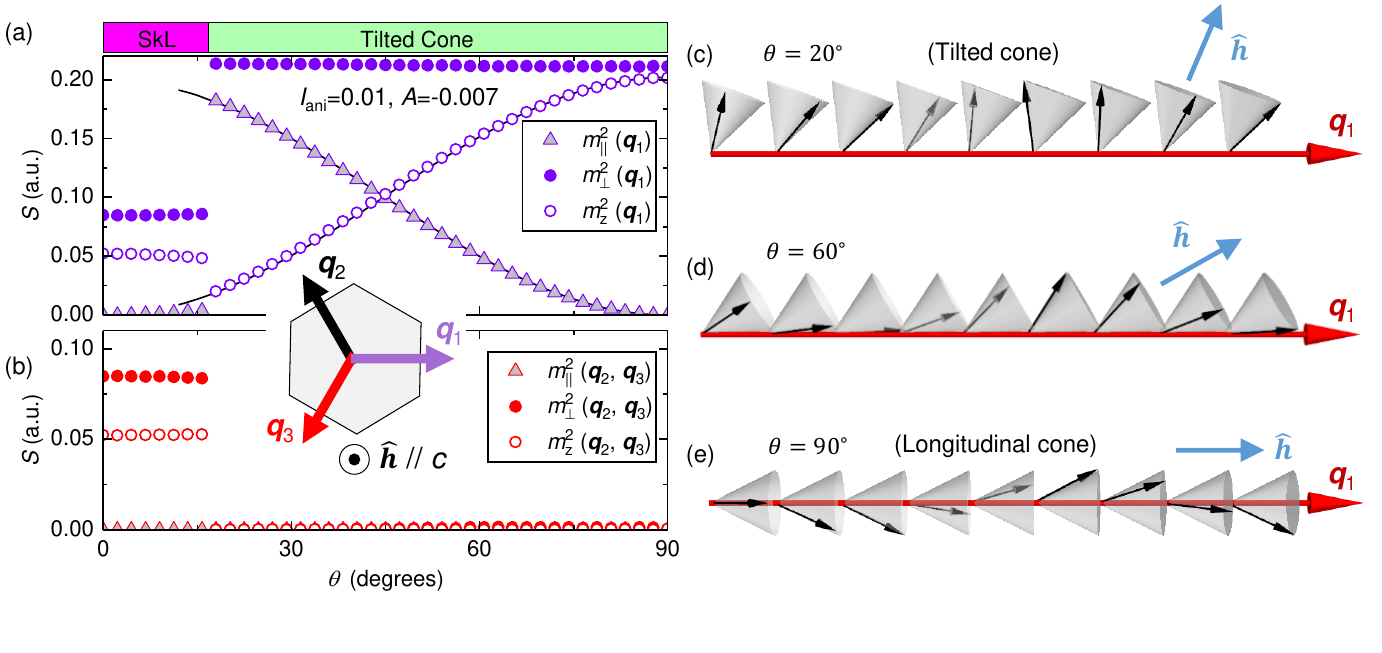}
		\internallinenumbers
    \caption[]{(color online). Modeling of tilting-induced destruction of the skyrmion lattice. (a,b) Components of the magnetic structure factor $S_{\mathbf{q}_\nu}$ for two reflections $\mathbf{q}_\nu$ at $h=0.8$ (c.f. Fig. 2). For this parameter set, the SkL collapses at tilt angle $\theta_c\approx 15^\circ$. In (a), the black solid lines show that $m_\parallel^2\sim \cos^2\theta$ and $m_z^2\sim \sin^2 \theta$ for $\theta>\theta_c$, respectively. (c-e) conical order at $\theta>\theta_c$, where the axis $\hat{\mathbf{n}}_1$ of the cone at $\mathbf{q}_1$ smoothly follows the direction of the magnetic field.}
    \label{fig:fig4}
  \end{center}
\end{figure}

\begin{figure}[!htb]
  \begin{center}
		\includegraphics[width=0.7\linewidth]{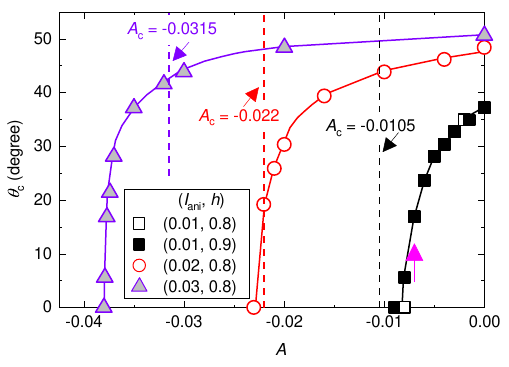}
		\internallinenumbers
    \caption[]{(color online). Critical angle $\theta_c$ of the skyrmion lattice and its dependence on model parameters. For each $I_\text{ani}=0.01$, $0.02$, $0.03$, a vertical line indicates the limit of stability for the helical (HE) state in zero magnetic field. When $A<A_c$, the $h=0$ ground state is transverse conical (TC). The parameters chosen for Gd$_3$Ru$_4$Al$_{12}$ are marked by a magenta arrow. For $I_\text{ani}=0.01$, two different values of the applied field $h$ -- normalized by exchange constant $J$ -- are used for the simulation, demonstrating the robustness of the result (solid and open squares). The lines passing through the data points are a guide to the eye.}
    \label{fig:fig5}
  \end{center}
\end{figure}

It is worth reiterating that the conical axis $\hat{\mathbf{n}}$ in Gd$_3$Ru$_4$Al$_{12}$ may be easily tilted away from being parallel to $\mathbf{q}$ (Figs. \ref{fig:fig4}, \ref{fig:fig5}). This experimental condition has potential applications in the field of reading and writing skyrmions by very weak external perturbations. Our study also indicates that depinning of the skyrmion lattice using an applied DC electrical current should be achievable in this material class. 

A recent focus of research on spiral magnets is the detection of the emergent electric field resulting from spin dynamics driven by electrical currents, i.e. the realization of an emergent (or quantum) inductor~\cite{Nagaosa2019}. In fact, the emergent inductance signal was recently observed by some of us in Gd$_3$Ru$_4$Al$_{12}$~\cite{Yokouchi2020}, exploring the zero-field helical (HE) phase -- discussed above -- as a prototypical spiral order of large local moments coupled to the Fermi sea. The present model indicates that the spiral plane in Gd$_3$Ru$_4$Al$_{12}$ is easily depinned and rotated, enabled by a balance of anisotropic exchange and local-ion anisotropy. This property facilitates spin dynamics, helping to make the emergent electric field observable in transport experiments. Beyond shedding light on the issue of skyrmion formation, we hope that our work can guide the ongoing search for room-temperature material hosts of the emergent inductance effect.

\textit{Acknowledgments.} We acknowledge discussions with J. Masell and T.-h. Arima. S. H. benefited from support by JSPS KAKENHI Grants Numbers JP18K13488, JP19K03752, and JP19H01834. This work was partially supported by JST CREST Grant Number JPMJCR1874 (Japan).

\bibliography{Bibliography}

\begin{thebibliography}{39}%
\makeatletter
\providecommand \@ifxundefined [1]{%
 \@ifx{#1\undefined}
}%
\providecommand \@ifnum [1]{%
 \ifnum #1\expandafter \@firstoftwo
 \else \expandafter \@secondoftwo
 \fi
}%
\providecommand \@ifx [1]{%
 \ifx #1\expandafter \@firstoftwo
 \else \expandafter \@secondoftwo
 \fi
}%
\providecommand \natexlab [1]{#1}%
\providecommand \enquote  [1]{``#1''}%
\providecommand \bibnamefont  [1]{#1}%
\providecommand \bibfnamefont [1]{#1}%
\providecommand \citenamefont [1]{#1}%
\providecommand \href@noop [0]{\@secondoftwo}%
\providecommand \href [0]{\begingroup \@sanitize@url \@href}%
\providecommand \@href[1]{\@@startlink{#1}\@@href}%
\providecommand \@@href[1]{\endgroup#1\@@endlink}%
\providecommand \@sanitize@url [0]{\catcode `\\12\catcode `\$12\catcode `\&12\catcode `\#12\catcode `\^12\catcode `\_12\catcode `\%12\relax}%
\providecommand \@@startlink[1]{}%
\providecommand \@@endlink[0]{}%
\providecommand \url  [0]{\begingroup\@sanitize@url \@url }%
\providecommand \@url [1]{\endgroup\@href {#1}{\urlprefix }}%
\providecommand \urlprefix  [0]{URL }%
\providecommand \Eprint [0]{\href }%
\providecommand \doibase [0]{http://dx.doi.org/}%
\providecommand \selectlanguage [0]{\@gobble}%
\providecommand \bibinfo  [0]{\@secondoftwo}%
\providecommand \bibfield  [0]{\@secondoftwo}%
\providecommand \translation [1]{[#1]}%
\providecommand \BibitemOpen [0]{}%
\providecommand \bibitemStop [0]{}%
\providecommand \bibitemNoStop [0]{.\EOS\space}%
\providecommand \EOS [0]{\spacefactor3000\relax}%
\providecommand \BibitemShut  [1]{\csname bibitem#1\endcsname}%
\let\auto@bib@innerbib\@empty
\bibitem [{\citenamefont {Okubo}\ \emph {et~al.}(2012)\citenamefont {Okubo}, \citenamefont {Chung},\ and\ \citenamefont {Kawamura}}]{Okubo2012}%
  \BibitemOpen
  \bibfield  {author} {\bibinfo {author} {\bibfnamefont {T.}~\bibnamefont {Okubo}}, \bibinfo {author} {\bibfnamefont {S.}~\bibnamefont {Chung}}, \ and\ \bibinfo {author} {\bibfnamefont {H.}~\bibnamefont {Kawamura}},\ }\bibfield  {title} {\enquote {\bibinfo {title} {Multiple-$q$ states and the skyrmion lattice of the triangular-lattice {H}eisenberg antiferromagnet under magnetic fields},}\ }\href {\doibase 10.1103/PhysRevLett.108.017206} {\bibfield  {journal} {\bibinfo  {journal} {Phys. Rev. Lett.}\ }\textbf {\bibinfo {volume} {108}},\ \bibinfo {pages} {017206} (\bibinfo {year} {2012})}\BibitemShut {NoStop}%
\bibitem [{\citenamefont {Leonov}\ and\ \citenamefont {Mostovoy}(2015)}]{Leonov2015}%
  \BibitemOpen
  \bibfield  {author} {\bibinfo {author} {\bibfnamefont {A.O.}\ \bibnamefont {Leonov}}\ and\ \bibinfo {author} {\bibfnamefont {M.}~\bibnamefont {Mostovoy}},\ }\bibfield  {title} {\enquote {\bibinfo {title} {Multiply periodic states and isolated skyrmions in an anisotropic frustrated magnet},}\ }\href {\doibase 10.1038/ncomms9275} {\bibfield  {journal} {\bibinfo  {journal} {Nature Communications}\ }\textbf {\bibinfo {volume} {6}},\ \bibinfo {pages} {8275} (\bibinfo {year} {2015})}\BibitemShut {NoStop}%
\bibitem [{\citenamefont {Ozawa}\ \emph {et~al.}(2017)\citenamefont {Ozawa}, \citenamefont {Hayami},\ and\ \citenamefont {Motome}}]{Ozawa2017}%
  \BibitemOpen
  \bibfield  {author} {\bibinfo {author} {\bibfnamefont {R.}~\bibnamefont {Ozawa}}, \bibinfo {author} {\bibfnamefont {S.}~\bibnamefont {Hayami}}, \ and\ \bibinfo {author} {\bibfnamefont {Y.}~\bibnamefont {Motome}},\ }\bibfield  {title} {\enquote {\bibinfo {title} {Zero-field skyrmions with a high topological number in itinerant magnets},}\ }\href@noop {} {\bibfield  {journal} {\bibinfo  {journal} {Physical Review Letters}\ }\textbf {\bibinfo {volume} {118}},\ \bibinfo {pages} {147205} (\bibinfo {year} {2017})}\BibitemShut {NoStop}%
\bibitem [{\citenamefont {Hayami}\ \emph {et~al.}(2017)\citenamefont {Hayami}, \citenamefont {Ozawa},\ and\ \citenamefont {Motome}}]{Hayami2017}%
  \BibitemOpen
  \bibfield  {author} {\bibinfo {author} {\bibfnamefont {S.}~\bibnamefont {Hayami}}, \bibinfo {author} {\bibfnamefont {R.}~\bibnamefont {Ozawa}}, \ and\ \bibinfo {author} {\bibfnamefont {Y.}~\bibnamefont {Motome}},\ }\bibfield  {title} {\enquote {\bibinfo {title} {Effective bilinear-biquadratic model for noncoplanar ordering in itinerant magnets},}\ }\href {\doibase 10.1103/PhysRevB.95.224424} {\bibfield  {journal} {\bibinfo  {journal} {Physical Review B}\ }\textbf {\bibinfo {volume} {95}},\ \bibinfo {pages} {224424} (\bibinfo {year} {2017})}\BibitemShut {NoStop}%
\bibitem [{\citenamefont {Wang}\ \emph {et~al.}(2020)\citenamefont {Wang}, \citenamefont {Su}, \citenamefont {Lin},\ and\ \citenamefont {Batista}}]{Wang2020}%
  \BibitemOpen
  \bibfield  {author} {\bibinfo {author} {\bibfnamefont {Z.}~\bibnamefont {Wang}}, \bibinfo {author} {\bibfnamefont {Y.}~\bibnamefont {Su}}, \bibinfo {author} {\bibfnamefont {S.-Z.}\ \bibnamefont {Lin}}, \ and\ \bibinfo {author} {\bibfnamefont {C.D.}\ \bibnamefont {Batista}},\ }\bibfield  {title} {\enquote {\bibinfo {title} {Skyrmion crystal from {RKKY} interaction mediated by 2{D} electron gas},}\ }\href@noop {} {\bibfield  {journal} {\bibinfo  {journal} {Physical Review Letters}\ }\textbf {\bibinfo {volume} {124}},\ \bibinfo {pages} {207201} (\bibinfo {year} {2020})}\BibitemShut {NoStop}%
\bibitem [{\citenamefont {Shimokawa}\ and\ \citenamefont {Kawamura}(2019)}]{Shimokawa2019}%
  \BibitemOpen
  \bibfield  {author} {\bibinfo {author} {\bibfnamefont {T.}~\bibnamefont {Shimokawa}}\ and\ \bibinfo {author} {\bibfnamefont {H.}~\bibnamefont {Kawamura}},\ }\bibfield  {title} {\enquote {\bibinfo {title} {{R}ipple state in the frustrated honeycomb-lattice antiferromagnet},}\ }\href {\doibase 10.1103/PhysRevLett.123.057202} {\bibfield  {journal} {\bibinfo  {journal} {Phys. Rev. Lett.}\ }\textbf {\bibinfo {volume} {123}},\ \bibinfo {pages} {057202} (\bibinfo {year} {2019})}\BibitemShut {NoStop}%
\bibitem [{\citenamefont {Lohani}\ \emph {et~al.}(2019)\citenamefont {Lohani}, \citenamefont {Hickey}, \citenamefont {Masell},\ and\ \citenamefont {Rosch}}]{Lohani2019}%
  \BibitemOpen
  \bibfield  {author} {\bibinfo {author} {\bibfnamefont {V.}~\bibnamefont {Lohani}}, \bibinfo {author} {\bibfnamefont {C.}~\bibnamefont {Hickey}}, \bibinfo {author} {\bibfnamefont {J.}~\bibnamefont {Masell}}, \ and\ \bibinfo {author} {\bibfnamefont {A.}~\bibnamefont {Rosch}},\ }\bibfield  {title} {\enquote {\bibinfo {title} {Quantum skyrmions in frustrated ferromagnets},}\ }\href@noop {} {\bibfield  {journal} {\bibinfo  {journal} {Physical Review X}\ }\textbf {\bibinfo {volume} {9}},\ \bibinfo {pages} {041063} (\bibinfo {year} {2019})}\BibitemShut {NoStop}%
\bibitem [{\citenamefont {Kurumaji}\ \emph {et~al.}(2019)\citenamefont {Kurumaji}, \citenamefont {Nakajima}, \citenamefont {Hirschberger}, \citenamefont {Kikkawa}, \citenamefont {Yamasaki}, \citenamefont {Sagayama}, \citenamefont {Nakao}, \citenamefont {Taguchi}, \citenamefont {Arima},\ and\ \citenamefont {Tokura}}]{Kurumaji2019}%
  \BibitemOpen
  \bibfield  {author} {\bibinfo {author} {\bibfnamefont {T.}~\bibnamefont {Kurumaji}}, \bibinfo {author} {\bibfnamefont {T.}~\bibnamefont {Nakajima}}, \bibinfo {author} {\bibfnamefont {M.}~\bibnamefont {Hirschberger}}, \bibinfo {author} {\bibfnamefont {A.}~\bibnamefont {Kikkawa}}, \bibinfo {author} {\bibfnamefont {Y.}~\bibnamefont {Yamasaki}}, \bibinfo {author} {\bibfnamefont {H.}~\bibnamefont {Sagayama}}, \bibinfo {author} {\bibfnamefont {H.}~\bibnamefont {Nakao}}, \bibinfo {author} {\bibfnamefont {Y.}~\bibnamefont {Taguchi}}, \bibinfo {author} {\bibfnamefont {T.-h.}\ \bibnamefont {Arima}}, \ and\ \bibinfo {author} {\bibfnamefont {Y.}~\bibnamefont {Tokura}},\ }\bibfield  {title} {\enquote {\bibinfo {title} {Skyrmion lattice with a giant topological {H}all effect in a frustrated triangular-lattice magnet},}\ }\href@noop {} {\bibfield  {journal} {\bibinfo  {journal} {Science}\ }\textbf {\bibinfo {volume} {365}},\ \bibinfo {pages} {914--918} (\bibinfo {year} {2019})}\BibitemShut {NoStop}%
\bibitem [{\citenamefont {Hirschberger}\ \emph {et~al.}(2019)\citenamefont {Hirschberger}, \citenamefont {Nakajima}, \citenamefont {Gao}, \citenamefont {Peng}, \citenamefont {Kikkawa}, \citenamefont {Kurumaji}, \citenamefont {Kriener}, \citenamefont {Yamasaki}, \citenamefont {Sagayama}, \citenamefont {Nakao}, \citenamefont {Ohishi}, \citenamefont {Kakurai}, \citenamefont {Taguchi}, \citenamefont {Yu}, \citenamefont {Arima},\ and\ \citenamefont {Tokura}}]{Hirschberger2019}%
  \BibitemOpen
  \bibfield  {author} {\bibinfo {author} {\bibfnamefont {M.}~\bibnamefont {Hirschberger}}, \bibinfo {author} {\bibfnamefont {T.}~\bibnamefont {Nakajima}}, \bibinfo {author} {\bibfnamefont {S.}~\bibnamefont {Gao}}, \bibinfo {author} {\bibfnamefont {L.}~\bibnamefont {Peng}}, \bibinfo {author} {\bibfnamefont {A.}~\bibnamefont {Kikkawa}}, \bibinfo {author} {\bibfnamefont {T.}~\bibnamefont {Kurumaji}}, \bibinfo {author} {\bibfnamefont {M.}~\bibnamefont {Kriener}}, \bibinfo {author} {\bibfnamefont {Y.}~\bibnamefont {Yamasaki}}, \bibinfo {author} {\bibfnamefont {H.}~\bibnamefont {Sagayama}}, \bibinfo {author} {\bibfnamefont {H.}~\bibnamefont {Nakao}}, \bibinfo {author} {\bibfnamefont {K.}~\bibnamefont {Ohishi}}, \bibinfo {author} {\bibfnamefont {K.}~\bibnamefont {Kakurai}}, \bibinfo {author} {\bibfnamefont {Y.}~\bibnamefont {Taguchi}}, \bibinfo {author} {\bibfnamefont {X.}~\bibnamefont {Yu}}, \bibinfo {author} {\bibfnamefont {T.-h.}\ \bibnamefont {Arima}}, \ and\ \bibinfo {author} {\bibfnamefont {Y.}~\bibnamefont
  {Tokura}},\ }\bibfield  {title} {\enquote {\bibinfo {title} {Skyrmion phase and competing magnetic orders on a breathing kagom{\'e} lattice},}\ }\href@noop {} {\bibfield  {journal} {\bibinfo  {journal} {Nature Communications}\ }\textbf {\bibinfo {volume} {10}},\ \bibinfo {pages} {5831} (\bibinfo {year} {2019})}\BibitemShut {NoStop}%
\bibitem [{\citenamefont {Gao}\ \emph {et~al.}(2017)\citenamefont {Gao}, \citenamefont {Zaharko}, \citenamefont {Tsurkan}, \citenamefont {Su}, \citenamefont {White}, \citenamefont {Tucker}, \citenamefont {Roessli}, \citenamefont {Bourdarot}, \citenamefont {Sibille}, \citenamefont {Chernyshov}, \citenamefont {Fennell}, \citenamefont {Loidl},\ and\ \citenamefont {Rüegg}}]{Gao2017}%
  \BibitemOpen
  \bibfield  {author} {\bibinfo {author} {\bibfnamefont {S.}~\bibnamefont {Gao}}, \bibinfo {author} {\bibfnamefont {O.}~\bibnamefont {Zaharko}}, \bibinfo {author} {\bibfnamefont {V.}~\bibnamefont {Tsurkan}}, \bibinfo {author} {\bibfnamefont {Y.}~\bibnamefont {Su}}, \bibinfo {author} {\bibfnamefont {J.S.}\ \bibnamefont {White}}, \bibinfo {author} {\bibfnamefont {G.S.}\ \bibnamefont {Tucker}}, \bibinfo {author} {\bibfnamefont {B.}~\bibnamefont {Roessli}}, \bibinfo {author} {\bibfnamefont {F.}~\bibnamefont {Bourdarot}}, \bibinfo {author} {\bibfnamefont {R.}~\bibnamefont {Sibille}}, \bibinfo {author} {\bibfnamefont {D.}~\bibnamefont {Chernyshov}}, \bibinfo {author} {\bibfnamefont {T.}~\bibnamefont {Fennell}}, \bibinfo {author} {\bibfnamefont {A.}~\bibnamefont {Loidl}}, \ and\ \bibinfo {author} {\bibfnamefont {C.}~\bibnamefont {Rüegg}},\ }\bibfield  {title} {\enquote {\bibinfo {title} {Spiral spin-liquid and the emergence of a vortex-like state in {M}n{S}c$_2${S}$_4$},}\ }\href@noop {} {\bibfield  {journal}
  {\bibinfo  {journal} {Nature Physics}\ }\textbf {\bibinfo {volume} {13}},\ \bibinfo {pages} {157--161} (\bibinfo {year} {2017})}\BibitemShut {NoStop}%
\bibitem [{\citenamefont {Gao}\ \emph {et~al.}(2020)\citenamefont {Gao}, \citenamefont {Rosales}, \citenamefont {Gómez~Albarracín}, \citenamefont {Tsurkan}, \citenamefont {Kaur}, \citenamefont {Fennell}, \citenamefont {Steffens}, \citenamefont {Boehm}, \citenamefont {Čermák}, \citenamefont {Schneidewind}, \citenamefont {Ressouche}, \citenamefont {Cabra}, \citenamefont {Rüegg},\ and\ \citenamefont {Zaharko}}]{Gao2020}%
  \BibitemOpen
  \bibfield  {author} {\bibinfo {author} {\bibfnamefont {S.}~\bibnamefont {Gao}}, \bibinfo {author} {\bibfnamefont {H.D.}\ \bibnamefont {Rosales}}, \bibinfo {author} {\bibfnamefont {F.A.}\ \bibnamefont {Gómez~Albarracín}}, \bibinfo {author} {\bibfnamefont {V.}~\bibnamefont {Tsurkan}}, \bibinfo {author} {\bibfnamefont {G.}~\bibnamefont {Kaur}}, \bibinfo {author} {\bibfnamefont {T.}~\bibnamefont {Fennell}}, \bibinfo {author} {\bibfnamefont {P.}~\bibnamefont {Steffens}}, \bibinfo {author} {\bibfnamefont {M.}~\bibnamefont {Boehm}}, \bibinfo {author} {\bibfnamefont {P.}~\bibnamefont {Čermák}}, \bibinfo {author} {\bibfnamefont {A.}~\bibnamefont {Schneidewind}}, \bibinfo {author} {\bibfnamefont {E.}~\bibnamefont {Ressouche}}, \bibinfo {author} {\bibfnamefont {D.C.}\ \bibnamefont {Cabra}}, \bibinfo {author} {\bibfnamefont {C.}~\bibnamefont {Rüegg}}, \ and\ \bibinfo {author} {\bibfnamefont {O.}~\bibnamefont {Zaharko}},\ }\bibfield  {title} {\enquote {\bibinfo {title} {Fractional antiferromagnetic skyrmion lattice
  induced by anisotropic couplings},}\ }\href@noop {} {\bibfield  {journal} {\bibinfo  {journal} {Nature}\ }\textbf {\bibinfo {volume} {586}},\ \bibinfo {pages} {37--41} (\bibinfo {year} {2020})}\BibitemShut {NoStop}%
\bibitem [{\citenamefont {Dzyaloshinskii}(1957)}]{Dzyaloshinskii1957}%
  \BibitemOpen
  \bibfield  {author} {\bibinfo {author} {\bibfnamefont {I.~E.}\ \bibnamefont {Dzyaloshinskii}},\ }\bibfield  {title} {\enquote {\bibinfo {title} {Thermodynamical theory of 'weak" ferromagnetism in antiferromagnetic substances},}\ }\href@noop {} {\bibfield  {journal} {\bibinfo  {journal} {Sov. Phys. JETP}\ }\textbf {\bibinfo {volume} {5}},\ \bibinfo {pages} {1259} (\bibinfo {year} {1957})}\BibitemShut {NoStop}%
\bibitem [{\citenamefont {Moriya}(1960)}]{Moriya1960}%
  \BibitemOpen
  \bibfield  {author} {\bibinfo {author} {\bibfnamefont {T\^oru}\ \bibnamefont {Moriya}},\ }\bibfield  {title} {\enquote {\bibinfo {title} {Anisotropic superexchange interaction and weak ferromagnetism},}\ }\href@noop {} {\bibfield  {journal} {\bibinfo  {journal} {Physical Review}\ }\textbf {\bibinfo {volume} {120}},\ \bibinfo {pages} {91--98} (\bibinfo {year} {1960})}\BibitemShut {NoStop}%
\bibitem [{\citenamefont {Bak}\ and\ \citenamefont {Jensen}(1980)}]{Bak1980}%
  \BibitemOpen
  \bibfield  {author} {\bibinfo {author} {\bibfnamefont {P.}~\bibnamefont {Bak}}\ and\ \bibinfo {author} {\bibfnamefont {M.H.}\ \bibnamefont {Jensen}},\ }\bibfield  {title} {\enquote {\bibinfo {title} {Theory of helical magnetic structures and phase transitions in {M}n{S}i and {F}e{G}e},}\ }\href@noop {} {\bibfield  {journal} {\bibinfo  {journal} {Journal of Physics C: Solid State Physics}\ }\textbf {\bibinfo {volume} {13}},\ \bibinfo {pages} {L881--L885} (\bibinfo {year} {1980})}\BibitemShut {NoStop}%
\bibitem [{\citenamefont {R{\"o}ßler}\ \emph {et~al.}(2006)\citenamefont {R{\"o}ßler}, \citenamefont {Bogdanov},\ and\ \citenamefont {Pfleiderer}}]{Roessler2006}%
  \BibitemOpen
  \bibfield  {author} {\bibinfo {author} {\bibfnamefont {U.~K.}\ \bibnamefont {R{\"o}ßler}}, \bibinfo {author} {\bibfnamefont {A.~N.}\ \bibnamefont {Bogdanov}}, \ and\ \bibinfo {author} {\bibfnamefont {C.}~\bibnamefont {Pfleiderer}},\ }\bibfield  {title} {\enquote {\bibinfo {title} {Spontaneous skyrmion ground states in magnetic metals},}\ }\href@noop {} {\bibfield  {journal} {\bibinfo  {journal} {Nature}\ }\textbf {\bibinfo {volume} {442}},\ \bibinfo {pages} {797--801} (\bibinfo {year} {2006})}\BibitemShut {NoStop}%
\bibitem [{\citenamefont {Kitaev}(2006)}]{Kitaev2006}%
  \BibitemOpen
  \bibfield  {author} {\bibinfo {author} {\bibfnamefont {A.}~\bibnamefont {Kitaev}},\ }\bibfield  {title} {\enquote {\bibinfo {title} {Anyons in an exactly solved model and beyond},}\ }\href@noop {} {\bibfield  {journal} {\bibinfo  {journal} {Annals of Physics}\ }\textbf {\bibinfo {volume} {321}},\ \bibinfo {pages} {2--111} (\bibinfo {year} {2006})}\BibitemShut {NoStop}%
\bibitem [{\citenamefont {Jackeli}\ and\ \citenamefont {Khaliullin}(2009)}]{Jackeli2009}%
  \BibitemOpen
  \bibfield  {author} {\bibinfo {author} {\bibfnamefont {G.}~\bibnamefont {Jackeli}}\ and\ \bibinfo {author} {\bibfnamefont {G.}~\bibnamefont {Khaliullin}},\ }\bibfield  {title} {\enquote {\bibinfo {title} {Mott insulators in the strong spin-orbit coupling limit: From {H}eisenberg to a quantum compass and {K}itaev models},}\ }\href@noop {} {\bibfield  {journal} {\bibinfo  {journal} {Physical Review Letters}\ }\textbf {\bibinfo {volume} {102}},\ \bibinfo {pages} {017205} (\bibinfo {year} {2009})}\BibitemShut {NoStop}%
\bibitem [{\citenamefont {Gladyshevskii}\ \emph {et~al.}(1993)\citenamefont {Gladyshevskii}, \citenamefont {Strusievicz}, \citenamefont {Cenzual},\ and\ \citenamefont {Parth{\'e}}}]{Gladyshevskii1993}%
  \BibitemOpen
  \bibfield  {author} {\bibinfo {author} {\bibfnamefont {R.~E.}\ \bibnamefont {Gladyshevskii}}, \bibinfo {author} {\bibfnamefont {O.~R.}\ \bibnamefont {Strusievicz}}, \bibinfo {author} {\bibfnamefont {K.}~\bibnamefont {Cenzual}}, \ and\ \bibinfo {author} {\bibfnamefont {E.}~\bibnamefont {Parth{\'e}}},\ }\bibfield  {title} {\enquote {\bibinfo {title} {Structure of {G}d$_3${R}u$_4${A}l$_{12}$, a new member of the {E}u{M}g$_{5.2}$ structure family with minority-atom clusters},}\ }\href@noop {} {\bibfield  {journal} {\bibinfo  {journal} {Acta Crystallographica Section B}\ }\textbf {\bibinfo {volume} {49}},\ \bibinfo {pages} {474--478} (\bibinfo {year} {1993})}\BibitemShut {NoStop}%
\bibitem [{\citenamefont {Niermann}\ and\ \citenamefont {Jeitschko}(2002)}]{Niermann2002}%
  \BibitemOpen
  \bibfield  {author} {\bibinfo {author} {\bibfnamefont {J.}~\bibnamefont {Niermann}}\ and\ \bibinfo {author} {\bibfnamefont {W.}~\bibnamefont {Jeitschko}},\ }\bibfield  {title} {\enquote {\bibinfo {title} {Ternary rare earth (${R}$) transition metal aluminides ${R}_3{T}_4${A}l$_{12}$ (${T}$ = {R}u and {O}s) with {G}d$_3${R}u$_4${A}l$_{12}$ type structure},}\ }\href@noop {} {\bibfield  {journal} {\bibinfo  {journal} {Journal for Inorganic and General Chemistry}\ }\textbf {\bibinfo {volume} {628}},\ \bibinfo {pages} {2549--2556} (\bibinfo {year} {2002})}\BibitemShut {NoStop}%
\bibitem [{\citenamefont {Chandragiri}\ \emph {et~al.}(2016)\citenamefont {Chandragiri}, \citenamefont {Iyer},\ and\ \citenamefont {Sampathkumaran}}]{Chandragiri2016}%
  \BibitemOpen
  \bibfield  {author} {\bibinfo {author} {\bibfnamefont {V.}~\bibnamefont {Chandragiri}}, \bibinfo {author} {\bibfnamefont {K.~K.}\ \bibnamefont {Iyer}}, \ and\ \bibinfo {author} {\bibfnamefont {E.V.}\ \bibnamefont {Sampathkumaran}},\ }\bibfield  {title} {\enquote {\bibinfo {title} {Magnetic behavior of {G}d$_3${R}u$_4${A}l$_{12}$, a layered compound with distorted kagom{\'e} net},}\ }\href@noop {} {\bibfield  {journal} {\bibinfo  {journal} {Journal of Physics: Condensed Matter}\ }\textbf {\bibinfo {volume} {28}},\ \bibinfo {pages} {286002} (\bibinfo {year} {2016})}\BibitemShut {NoStop}%
\bibitem [{\citenamefont {Matsumura}\ \emph {et~al.}(2019)\citenamefont {Matsumura}, \citenamefont {Ozono}, \citenamefont {Nakamura}, \citenamefont {Kabeya},\ and\ \citenamefont {Ochiai}}]{Matsumura2019}%
  \BibitemOpen
  \bibfield  {author} {\bibinfo {author} {\bibfnamefont {T.}~\bibnamefont {Matsumura}}, \bibinfo {author} {\bibfnamefont {Y.}~\bibnamefont {Ozono}}, \bibinfo {author} {\bibfnamefont {S.}~\bibnamefont {Nakamura}}, \bibinfo {author} {\bibfnamefont {N.}~\bibnamefont {Kabeya}}, \ and\ \bibinfo {author} {\bibfnamefont {A.}~\bibnamefont {Ochiai}},\ }\bibfield  {title} {\enquote {\bibinfo {title} {Helical ordering of spin trimers in a distorted {K}agome lattice of {G}d$_3${R}u$_4${A}l$_{12}$ studied by resonant x-ray diffraction},}\ }\href@noop {} {\bibfield  {journal} {\bibinfo  {journal} {J. Phys. Soc. Jpn.}\ }\textbf {\bibinfo {volume} {88}},\ \bibinfo {pages} {023704} (\bibinfo {year} {2019})}\BibitemShut {NoStop}%
\bibitem [{\citenamefont {Nakamura}\ \emph {et~al.}(2018)\citenamefont {Nakamura}, \citenamefont {Kabeya}, \citenamefont {Kobayashi}, \citenamefont {Araki}, \citenamefont {Katoh},\ and\ \citenamefont {Ochiai}}]{Nakamura2018}%
  \BibitemOpen
  \bibfield  {author} {\bibinfo {author} {\bibfnamefont {S.}~\bibnamefont {Nakamura}}, \bibinfo {author} {\bibfnamefont {N.}~\bibnamefont {Kabeya}}, \bibinfo {author} {\bibfnamefont {M.}~\bibnamefont {Kobayashi}}, \bibinfo {author} {\bibfnamefont {K.}~\bibnamefont {Araki}}, \bibinfo {author} {\bibfnamefont {K.}~\bibnamefont {Katoh}}, \ and\ \bibinfo {author} {\bibfnamefont {A.}~\bibnamefont {Ochiai}},\ }\bibfield  {title} {\enquote {\bibinfo {title} {Spin trimer formation in the metallic compound {G}d$_3${R}u$_4${A}l$_{12}$ with a distorted kagome lattice structure},}\ }\href@noop {} {\bibfield  {journal} {\bibinfo  {journal} {Physical Review B}\ }\textbf {\bibinfo {volume} {98}},\ \bibinfo {pages} {054410} (\bibinfo {year} {2018})}\BibitemShut {NoStop}%
\bibitem [{\citenamefont {Li}\ \emph {et~al.}(2016)\citenamefont {Li}, \citenamefont {Wang},\ and\ \citenamefont {Chen}}]{Li2016}%
  \BibitemOpen
  \bibfield  {author} {\bibinfo {author} {\bibfnamefont {Y.-D.}\ \bibnamefont {Li}}, \bibinfo {author} {\bibfnamefont {X.}~\bibnamefont {Wang}}, \ and\ \bibinfo {author} {\bibfnamefont {G.}~\bibnamefont {Chen}},\ }\bibfield  {title} {\enquote {\bibinfo {title} {Anisotropic spin model of strong spin-orbit-coupled triangular antiferromagnets},}\ }\href@noop {} {\bibfield  {journal} {\bibinfo  {journal} {Physical Review B}\ }\textbf {\bibinfo {volume} {94}},\ \bibinfo {pages} {035107} (\bibinfo {year} {2016})}\BibitemShut {NoStop}%
\bibitem [{\citenamefont {Hayami}\ and\ \citenamefont {Motome}(2018)}]{Hayami2018}%
  \BibitemOpen
  \bibfield  {author} {\bibinfo {author} {\bibfnamefont {S.}~\bibnamefont {Hayami}}\ and\ \bibinfo {author} {\bibfnamefont {Y.}~\bibnamefont {Motome}},\ }\bibfield  {title} {\enquote {\bibinfo {title} {N{\'e}el- and {B}loch-type magnetic vortices in {R}ashba metals},}\ }\href@noop {} {\bibfield  {journal} {\bibinfo  {journal} {Physical Review Letters}\ }\textbf {\bibinfo {volume} {121}},\ \bibinfo {pages} {137202} (\bibinfo {year} {2018})}\BibitemShut {NoStop}%
\bibitem [{\citenamefont {Rybakov}\ \emph {et~al.}(2016)\citenamefont {Rybakov}, \citenamefont {Borisov}, \citenamefont {Bl{\"u}gel},\ and\ \citenamefont {Kiselev}}]{Rybakov2016}%
  \BibitemOpen
  \bibfield  {author} {\bibinfo {author} {\bibfnamefont {F.N.}\ \bibnamefont {Rybakov}}, \bibinfo {author} {\bibfnamefont {A.B.}\ \bibnamefont {Borisov}}, \bibinfo {author} {\bibfnamefont {S.}~\bibnamefont {Bl{\"u}gel}}, \ and\ \bibinfo {author} {\bibfnamefont {N.S.}\ \bibnamefont {Kiselev}},\ }\bibfield  {title} {\enquote {\bibinfo {title} {New spiral state and skyrmion lattice in 3{D} model of chiral magnets},}\ }\href@noop {} {\bibfield  {journal} {\bibinfo  {journal} {New Journal of Physics}\ }\textbf {\bibinfo {volume} {18}},\ \bibinfo {pages} {045002} (\bibinfo {year} {2016})}\BibitemShut {NoStop}%
\bibitem [{\citenamefont {Puphal}\ \emph {et~al.}(2020)\citenamefont {Puphal}, \citenamefont {Pomjakushin}, \citenamefont {Kanazawa}, \citenamefont {Ukleev}, \citenamefont {Gawryluk}, \citenamefont {Ma}, \citenamefont {Naamneh}, \citenamefont {Plumb}, \citenamefont {Keller}, \citenamefont {Cubitt}, \citenamefont {Pomjakushina},\ and\ \citenamefont {White}}]{Puphal2020}%
  \BibitemOpen
  \bibfield  {author} {\bibinfo {author} {\bibfnamefont {P.}~\bibnamefont {Puphal}}, \bibinfo {author} {\bibfnamefont {V.}~\bibnamefont {Pomjakushin}}, \bibinfo {author} {\bibfnamefont {N.}~\bibnamefont {Kanazawa}}, \bibinfo {author} {\bibfnamefont {V.}~\bibnamefont {Ukleev}}, \bibinfo {author} {\bibfnamefont {D.~J.}\ \bibnamefont {Gawryluk}}, \bibinfo {author} {\bibfnamefont {J.}~\bibnamefont {Ma}}, \bibinfo {author} {\bibfnamefont {M.}~\bibnamefont {Naamneh}}, \bibinfo {author} {\bibfnamefont {N.~C.}\ \bibnamefont {Plumb}}, \bibinfo {author} {\bibfnamefont {L.}~\bibnamefont {Keller}}, \bibinfo {author} {\bibfnamefont {R.}~\bibnamefont {Cubitt}}, \bibinfo {author} {\bibfnamefont {E.}~\bibnamefont {Pomjakushina}}, \ and\ \bibinfo {author} {\bibfnamefont {J.~S.}\ \bibnamefont {White}},\ }\bibfield  {title} {\enquote {\bibinfo {title} {Topological magnetic phase in the candidate weyl semimetal cealges},}\ }\href@noop {} {\bibfield  {journal} {\bibinfo  {journal} {Physical Review Letters}\ }\textbf {\bibinfo
  {volume} {124}},\ \bibinfo {pages} {017202} (\bibinfo {year} {2020})}\BibitemShut {NoStop}%
\bibitem [{\citenamefont {Nomoto}\ \emph {et~al.}(2020)\citenamefont {Nomoto}, \citenamefont {Koretsune},\ and\ \citenamefont {Arita}}]{Nomoto2020}%
  \BibitemOpen
  \bibfield  {author} {\bibinfo {author} {\bibfnamefont {T.}~\bibnamefont {Nomoto}}, \bibinfo {author} {\bibfnamefont {T.}~\bibnamefont {Koretsune}}, \ and\ \bibinfo {author} {\bibfnamefont {R.}~\bibnamefont {Arita}},\ }\bibfield  {title} {\enquote {\bibinfo {title} {Formation mechanism of the helical $\mathbf{Q}$ structure in {G}d-based skyrmion materials},}\ }\href@noop {} {\bibfield  {journal} {\bibinfo  {journal} {Phys. Rev. Lett.}\ }\textbf {\bibinfo {volume} {125}},\ \bibinfo {pages} {117204} (\bibinfo {year} {2020})}\BibitemShut {NoStop}%
\bibitem [{SI()}]{SI}%
  \BibitemOpen
  \href@noop {} {\emph {\bibinfo {title} {Supplementary Information}}}\BibitemShut {NoStop}%
\bibitem [{\citenamefont {Bruno}\ \emph {et~al.}(2004)\citenamefont {Bruno}, \citenamefont {Dugaev},\ and\ \citenamefont {Taillefumier}}]{Bruno2004}%
  \BibitemOpen
  \bibfield  {author} {\bibinfo {author} {\bibfnamefont {P.}~\bibnamefont {Bruno}}, \bibinfo {author} {\bibfnamefont {V.K.}\ \bibnamefont {Dugaev}}, \ and\ \bibinfo {author} {\bibfnamefont {M.}~\bibnamefont {Taillefumier}},\ }\bibfield  {title} {\enquote {\bibinfo {title} {Topological {H}all effect and {B}erry phase in magnetic nanostructures},}\ }\href {\doibase 10.1103/PhysRevLett.93.096806} {\bibfield  {journal} {\bibinfo  {journal} {Phys. Rev. Lett.}\ }\textbf {\bibinfo {volume} {93}},\ \bibinfo {pages} {096806} (\bibinfo {year} {2004})}\BibitemShut {NoStop}%
\bibitem [{\citenamefont {Neubauer}\ \emph {et~al.}(2009)\citenamefont {Neubauer}, \citenamefont {Pfleiderer}, \citenamefont {Binz}, \citenamefont {Rosch}, \citenamefont {Ritz}, \citenamefont {Niklowitz},\ and\ \citenamefont {B{\"o}ni}}]{Neubauer2009}%
  \BibitemOpen
  \bibfield  {author} {\bibinfo {author} {\bibfnamefont {A.}~\bibnamefont {Neubauer}}, \bibinfo {author} {\bibfnamefont {C.}~\bibnamefont {Pfleiderer}}, \bibinfo {author} {\bibfnamefont {B.}~\bibnamefont {Binz}}, \bibinfo {author} {\bibfnamefont {A.}~\bibnamefont {Rosch}}, \bibinfo {author} {\bibfnamefont {R.}~\bibnamefont {Ritz}}, \bibinfo {author} {\bibfnamefont {P.G.}\ \bibnamefont {Niklowitz}}, \ and\ \bibinfo {author} {\bibfnamefont {P.}~\bibnamefont {B{\"o}ni}},\ }\bibfield  {title} {\enquote {\bibinfo {title} {Topological {H}all effect in the ${A}$ phase of {M}n{S}i},}\ }\href {\doibase 10.1103/PhysRevLett.102.186602} {\bibfield  {journal} {\bibinfo  {journal} {Phys. Rev. Lett.}\ }\textbf {\bibinfo {volume} {102}},\ \bibinfo {pages} {186602} (\bibinfo {year} {2009})}\BibitemShut {NoStop}%
\bibitem [{\citenamefont {Ritz}\ \emph {et~al.}(2013)\citenamefont {Ritz}, \citenamefont {Halder}, \citenamefont {Franz}, \citenamefont {Bauer}, \citenamefont {Wagner}, \citenamefont {Bamler}, \citenamefont {Rosch},\ and\ \citenamefont {Pfleiderer}}]{Ritz2013}%
  \BibitemOpen
  \bibfield  {author} {\bibinfo {author} {\bibfnamefont {R.}~\bibnamefont {Ritz}}, \bibinfo {author} {\bibfnamefont {M.}~\bibnamefont {Halder}}, \bibinfo {author} {\bibfnamefont {C.}~\bibnamefont {Franz}}, \bibinfo {author} {\bibfnamefont {A.}~\bibnamefont {Bauer}}, \bibinfo {author} {\bibfnamefont {M.}~\bibnamefont {Wagner}}, \bibinfo {author} {\bibfnamefont {R.}~\bibnamefont {Bamler}}, \bibinfo {author} {\bibfnamefont {A.}~\bibnamefont {Rosch}}, \ and\ \bibinfo {author} {\bibfnamefont {C.}~\bibnamefont {Pfleiderer}},\ }\bibfield  {title} {\enquote {\bibinfo {title} {Giant generic topological {H}all resistivity of {M}n{S}i under pressure},}\ }\href {\doibase 10.1103/PhysRevB.87.134424} {\bibfield  {journal} {\bibinfo  {journal} {Phys. Rev. B}\ }\textbf {\bibinfo {volume} {87}},\ \bibinfo {pages} {134424} (\bibinfo {year} {2013})}\BibitemShut {NoStop}%
\bibitem [{\citenamefont {Nagaosa}\ \emph {et~al.}(2010)\citenamefont {Nagaosa}, \citenamefont {Sinova}, \citenamefont {Onoda}, \citenamefont {MacDonald},\ and\ \citenamefont {Ong}}]{Nagaosa2010}%
  \BibitemOpen
  \bibfield  {author} {\bibinfo {author} {\bibfnamefont {N.}~\bibnamefont {Nagaosa}}, \bibinfo {author} {\bibfnamefont {J.}~\bibnamefont {Sinova}}, \bibinfo {author} {\bibfnamefont {S.}~\bibnamefont {Onoda}}, \bibinfo {author} {\bibfnamefont {A.H.}\ \bibnamefont {MacDonald}}, \ and\ \bibinfo {author} {\bibfnamefont {N.P.}\ \bibnamefont {Ong}},\ }\bibfield  {title} {\enquote {\bibinfo {title} {Anomalous {H}all effect},}\ }\href {\doibase 10.1103/RevModPhys.82.1539} {\bibfield  {journal} {\bibinfo  {journal} {Rev. Mod. Phys.}\ }\textbf {\bibinfo {volume} {82}},\ \bibinfo {pages} {1539--1592} (\bibinfo {year} {2010})}\BibitemShut {NoStop}%
\bibitem [{\citenamefont {Karplus}\ and\ \citenamefont {Luttinger}(1954)}]{Karplus1954}%
  \BibitemOpen
  \bibfield  {author} {\bibinfo {author} {\bibfnamefont {R.}~\bibnamefont {Karplus}}\ and\ \bibinfo {author} {\bibfnamefont {J.~M.}\ \bibnamefont {Luttinger}},\ }\bibfield  {title} {\enquote {\bibinfo {title} {Hall effect in ferromagnetics},}\ }\href@noop {} {\bibfield  {journal} {\bibinfo  {journal} {Physical Review}\ }\textbf {\bibinfo {volume} {95}},\ \bibinfo {pages} {1154} (\bibinfo {year} {1954})}\BibitemShut {NoStop}%
\bibitem [{\citenamefont {Lee}\ \emph {et~al.}(2007)\citenamefont {Lee}, \citenamefont {Onose}, \citenamefont {Tokura},\ and\ \citenamefont {Ong}}]{Lee2007}%
  \BibitemOpen
  \bibfield  {author} {\bibinfo {author} {\bibfnamefont {M.}~\bibnamefont {Lee}}, \bibinfo {author} {\bibfnamefont {Y.}~\bibnamefont {Onose}}, \bibinfo {author} {\bibfnamefont {Y.}~\bibnamefont {Tokura}}, \ and\ \bibinfo {author} {\bibfnamefont {N.~P.}\ \bibnamefont {Ong}},\ }\bibfield  {title} {\enquote {\bibinfo {title} {Hidden constant in the anomalous hall effect of high-purity magnet mnsi},}\ }\href@noop {} {\bibfield  {journal} {\bibinfo  {journal} {Phys. Rev. B}\ }\textbf {\bibinfo {volume} {75}},\ \bibinfo {pages} {172403} (\bibinfo {year} {2007})}\BibitemShut {NoStop}%
\bibitem [{\citenamefont {Hayami}\ \emph {et~al.}(2016)\citenamefont {Hayami}, \citenamefont {Lin},\ and\ \citenamefont {Batista}}]{Hayami2016}%
  \BibitemOpen
  \bibfield  {author} {\bibinfo {author} {\bibfnamefont {S.}~\bibnamefont {Hayami}}, \bibinfo {author} {\bibfnamefont {S.-Z.}\ \bibnamefont {Lin}}, \ and\ \bibinfo {author} {\bibfnamefont {C.D.}\ \bibnamefont {Batista}},\ }\bibfield  {title} {\enquote {\bibinfo {title} {Bubble and skyrmion crystals in frustrated magnets with easy-axis anisotropy},}\ }\href {\doibase 10.1103/PhysRevB.93.184413} {\bibfield  {journal} {\bibinfo  {journal} {Phys. Rev. B}\ }\textbf {\bibinfo {volume} {93}},\ \bibinfo {pages} {184413} (\bibinfo {year} {2016})}\BibitemShut {NoStop}%
\bibitem [{\citenamefont {Bord{\'a}cs}\ \emph {et~al.}(2017)\citenamefont {Bord{\'a}cs}, \citenamefont {Butykai}, \citenamefont {Szigeti}, \citenamefont {White}, \citenamefont {Cubitt}, \citenamefont {Leonov}, \citenamefont {Widmann}, \citenamefont {Ehlers}, \citenamefont {von Nidda}, \citenamefont {Tsurkan}, \citenamefont {Loidl},\ and\ \citenamefont {K{\'e}zsm{\'a}rki}}]{Bordacs2017}%
  \BibitemOpen
  \bibfield  {author} {\bibinfo {author} {\bibfnamefont {S.}~\bibnamefont {Bord{\'a}cs}}, \bibinfo {author} {\bibfnamefont {A.}~\bibnamefont {Butykai}}, \bibinfo {author} {\bibfnamefont {B.~G.}\ \bibnamefont {Szigeti}}, \bibinfo {author} {\bibfnamefont {J.~S.}\ \bibnamefont {White}}, \bibinfo {author} {\bibfnamefont {R.}~\bibnamefont {Cubitt}}, \bibinfo {author} {\bibfnamefont {A.~O.}\ \bibnamefont {Leonov}}, \bibinfo {author} {\bibfnamefont {S.}~\bibnamefont {Widmann}}, \bibinfo {author} {\bibfnamefont {D.}~\bibnamefont {Ehlers}}, \bibinfo {author} {\bibfnamefont {H.-A.~Krug}\ \bibnamefont {von Nidda}}, \bibinfo {author} {\bibfnamefont {V.}~\bibnamefont {Tsurkan}}, \bibinfo {author} {\bibfnamefont {A.}~\bibnamefont {Loidl}}, \ and\ \bibinfo {author} {\bibfnamefont {I.}~\bibnamefont {K{\'e}zsm{\'a}rki}},\ }\bibfield  {title} {\enquote {\bibinfo {title} {Equilibrium skyrmion lattice ground state in a polar easy-plane magnet},}\ }\href@noop {} {\bibfield  {journal} {\bibinfo  {journal} {Scientific Reports}\
  }\textbf {\bibinfo {volume} {7}},\ \bibinfo {pages} {7584} (\bibinfo {year} {2017})}\BibitemShut {NoStop}%
\bibitem [{\citenamefont {Leonov}\ and\ \citenamefont {K{\'e}zsm{\'a}rki}(2017)}]{Leonov2017}%
  \BibitemOpen
  \bibfield  {author} {\bibinfo {author} {\bibfnamefont {A.~O.}\ \bibnamefont {Leonov}}\ and\ \bibinfo {author} {\bibfnamefont {I.}~\bibnamefont {K{\'e}zsm{\'a}rki}},\ }\bibfield  {title} {\enquote {\bibinfo {title} {Skyrmion robustness in noncentrosymmetric magnets with axial symmetry: The role of anisotropy and tilted magnetic fields},}\ }\href@noop {} {\bibfield  {journal} {\bibinfo  {journal} {Physical Review B}\ }\textbf {\bibinfo {volume} {96}},\ \bibinfo {pages} {214413} (\bibinfo {year} {2017})}\BibitemShut {NoStop}%
\bibitem [{\citenamefont {Nagaosa}(2019)}]{Nagaosa2019}%
  \BibitemOpen
  \bibfield  {author} {\bibinfo {author} {\bibfnamefont {N.}~\bibnamefont {Nagaosa}},\ }\bibfield  {title} {\enquote {\bibinfo {title} {Emergent inductor by spiral magnets},}\ }\href@noop {} {\bibfield  {journal} {\bibinfo  {journal} {Jpn. J. Appl. Phys.}\ }\textbf {\bibinfo {volume} {58}},\ \bibinfo {pages} {12} (\bibinfo {year} {2019})}\BibitemShut {NoStop}%
\bibitem [{\citenamefont {Yokouchi}\ \emph {et~al.}(2020)\citenamefont {Yokouchi}, \citenamefont {Kagawa}, \citenamefont {Hirschberger}, \citenamefont {Otani}, \citenamefont {Nagaosa},\ and\ \citenamefont {Tokura}}]{Yokouchi2020}%
  \BibitemOpen
  \bibfield  {author} {\bibinfo {author} {\bibfnamefont {T.}~\bibnamefont {Yokouchi}}, \bibinfo {author} {\bibfnamefont {F.}~\bibnamefont {Kagawa}}, \bibinfo {author} {\bibfnamefont {M.}~\bibnamefont {Hirschberger}}, \bibinfo {author} {\bibfnamefont {Y.}~\bibnamefont {Otani}}, \bibinfo {author} {\bibfnamefont {N.}~\bibnamefont {Nagaosa}}, \ and\ \bibinfo {author} {\bibfnamefont {Y.}~\bibnamefont {Tokura}},\ }\bibfield  {title} {\enquote {\bibinfo {title} {Emergent electromagnetic induction in a helical-spin magnet},}\ }\href@noop {} {\bibfield  {journal} {\bibinfo  {journal} {Nature}\ }\textbf {\bibinfo {volume} {586}},\ \bibinfo {pages} {232–236} (\bibinfo {year} {2020})}\BibitemShut {NoStop}%
\end{thebibliography}%

\clearpage

\end{document}